\documentclass{PoS}

\usepackage[authoryear,square]{natbib}
\bibpunct{(}{)}{;}{a}{}{,}

\title{Unravelling lifecycles \& physics of radio-loud AGN in the SKA era}

\ShortTitle{The lifecycles \& physics of radio-loud AGN}

\author{
  \speaker{Anna D. Kapi\'{n}ska}$^{,1,2}$, 
  Martin J. Hardcastle$^{*,3}$, 
  Carole A. Jackson$^{4}$,
  Tao An$^{5,6}$,
  Willem A. Baan$^5$,
  Matt J. Jarvis$^{7,8}$
  \\ 
  $^1$ ARC Centre of Excellence for All-Sky Astrophysics (CAASTRO);
  $^2$ ICRAR, The University of Western Australia, M468, 35 Stirling Hwy, Crawley WA 6009, Australia;
  $^3$ School of Physics, Astronomy and Mathematics, University of Hertfordshire, College Lane, Hatfield AL10 9AB, United Kingdom;
  $^4$ ICRAR, Curtin University, GPO Box U1987, Perth WA 6845, Australia;\\
  $^5$ Shanghai Astronomical Observatory, Chinese Academy of Sciences, 200030 Shanghai, China;\\
  $^6$ Key Laboratory of Radio Astronomy, Chinese Academy of Sciences, 210008 Nanjing, China;\\
  $^7$ Oxford Astrophysics, Department of Physics, Keble Road, Oxford OX1 3RH, United Kingdom; \\
  $^8$ Department of Physics, University of Western Cape, Cape Town 7535, South Africa
  \\
  E-mail: \email{anna.kapinska at uwa.edu.au}
}

\abstract{
Radio-loud AGN ($>10^{22}$~W~Hz$^{-1}$ at 1.4~GHz) will be the dominant bright source population detected with the SKA. The high resolution that the SKA will provide even in wide-area surveys will mean that, for the first time sensitive, multi-frequency total intensity and polarisation imaging of large samples of radio-loud active galactic nuclei (AGN) will become available. The unprecedented sensitivity of the SKA coupled with its wide field of view capabilities will allow identification of objects of the same morphological type (i.e. the entire FR~I, low- and high-luminosity FR~II, disturbed morphology  as well as weak radio-emitting AGN populations) up to high redshifts ($z\sim4$ and beyond), and at the same stage of their lives, from the youngest CSS/GPS sources to giant and fading (dying) sources, through to those with restarted activity radio galaxies and quasars. Critically, the wide frequency coverage of the SKA will permit analysis of same-epoch rest-frame radio properties, and the sensitivity and resolution will allow full cross-identification with multi-waveband data, further revealing insights into the physical processes driving the evolution of these radio sources. In this chapter of the SKA Science Book we give a summary of the main science drivers in the studies of lifecycles and detailed physics of radio-loud AGN, which include radio and kinetic luminosity functions, AGN feedback, radio-AGN triggering, radio-loud AGN unification and cosmological studies. We discuss the best parameters for the proposed SKA continuum surveys, both all-sky and deep field, in the light of these studies.
}

\FullConference{
Advancing Astrophysics with the Square Kilometre Array\\
June 8-13, 2014\\
Giardini Naxos, Sicily, Italy}

\newcommand{\skipthis}[1]{}
\newcommand{\apj}{ApJ}
\newcommand{\apjl}{ApJL}

\newcommand{\aj}{AJ}
\newcommand{\apl}{ApL}
\newcommand{\aap}{A\&A}

\newcommand{\mnras}{MNRAS}
\newcommand{\nar}{NewAR}
\newcommand{\nat}{Nature}

\newcommand{\pasp}{PASP}
\newcommand{\pasa}{PASA}

\newcommand{\araa}{ARA\&A}
\newcommand{\etal}{et al}
\newcommand{\an}{AN}

\begin{document}

\section{Introduction}
\label{sec:intro}

Large-scale radio surveys compiled over the past 50 years have revealed a multitude of source types that we term `radio-loud AGN' \cite[][]{1982MNRAS.198..843P,1994AuJPh..47..625W}. These objects will be the dominant population of bright sources detected by the SKA. The deep SKA radio surveys will allow detailed and complete analysis of most, if not all, radio-loud AGN populations and provide the basis to resolve some of the most critical questions in this area.

The radio emission from radio-loud AGN is synchrotron emission produced by a population of electrons, transported in a relativistic outflow from the vicinity of the central supermassive black hole, and accelerated to high energies as the jet expands and decelerates to sub-relativistic speeds. This is the dominating source of radio emission from radio-loud AGN down to luminosity densities of $\sim10^{22}$~W~Hz$^{-1}$ at 1.4~GHz; below this luminosity density AGN are referred to as weakly radio-emitting and it is under debate whether their radio emission is still dominated by an active nucleus rather than star formation. The subject of weakly radio-emitting AGN is covered elsewhere in this volume \citep[][see also \S\ref{sec:quiet}]{AASKAOrienti}. Based on unification models \cite[e.g.][]{1989ApJ...336..606B,1995PASP..107..803U,1999MNRAS.304..160J}, radio-loud AGN are often distinguished as radio galaxies, radio-loud quasars and BL~Lac-type objects (dependent on the orientation to the observer). Radio galaxies and radio-loud quasars are further broadly divided into unresolved or compact symmetric objects, FR~I and FR~II type sources (including peculiar morphologies) and giant radio galaxies (see \S\ref{sec:picture}; classification of different life stages and radio morphologies). Here, we consider all of these radio-loud AGN, with no selection on multi-wavelength properties of their hosts.

The physics of, and physical conditions in, the radio-emitting plasma are of great interest in themselves, but radio-loud AGN are important for a number of other reasons. Firstly, they provide an obscuration-independent method of selecting AGN out to the highest redshifts, and in some cases, e.g. radiatively inefficient, low accretion-rate AGN radio emission gives us the {\it only} method of measuring the output and accretion rate of the system. Secondly, radio-loud AGN are now routinely invoked in models of galaxy formation and evolution, where they provide a so-called `feedback' mechanism. AGN feedback is now thought to be one of the main mechanisms preventing the cooling of large-scale gas and the consequent growth of the host galaxies \cite[][]{2006MNRAS.370..645B,2006MNRAS.365...11C}. Finally, we expect the SKA surveys to reach nJy levels, detecting statistically significant numbers of sources across the wide range of the radio luminosity function (RLF) \emph{at all cosmic epochs}. When coupled with sufficient angular resolution and multi-waveband data, it will be possible to separate the contribution of radio emission due to an active nucleus from that due to ongoing or bursting star formation \cite[SF;][]{AASKAMcAlpine}.

In this Chapter we primarily focus on how the SKA will reveal the evolution of the radio-loud AGN populations characterised by their radio morphologies and luminosity densities, and at the same time directly provide the necessary radio data for studies of the radio source physics for the first time. There are a number of key questions that these deep samples can address.

\begin{itemize}

\item What is the RLF at all cosmic epochs? 

There is a huge range of radio AGN luminosity densities; in the local Universe this extends from $10^{22}$ to $10^{27}$ W Hz$^{-1}$ at 1.4 GHz \cite[][]{2007MNRAS.375..931M}. Due to the Malmquist effect, deep samples are highly biased towards high luminosity sources near to the limiting magnitude at each epoch. The result is that RLFs derived from `complete' small-area radio samples are limited in accuracy and fail to fully probe the breadth of the full RLF. Although it is well established that the RLF evolves steeply for the overall radio-loud AGN population \cite[note the steepness is luminosity dependent, e.g.][among many others]{1990MNRAS.247...19D, 2007MNRAS.381..211S, 2009MNRAS.392..617D, 2011MNRAS.416.1900R, 2013MNRAS.436.1084M, 2014MNRAS.445..955B}, it is still not clear whether it is sources themselves or number of sources that become brighter.

\item Is there a link between the evolving radio-loud AGN RLF and the evolution of galaxies and galaxy clusters? Are feedback processes inherent to all radio-loud AGN, or to just a subset of them, and what does this reveal about the physical processes within these populations?  

Although interaction between the radio lobes and the hot ambient medium is directly observable in X-rays in the local Universe, it is still an open question whether the physics of radio galaxies is consistent with the role they are thought to play in the models of galaxy formation and evolution \cite[][]{2009Natur.460..213C,2012NJPh...14e5023M}. Studies of radio-loud AGN populations in the local Universe \cite[e.g.][]{2012MNRAS.421.1569B} show that there is a fundamental dichotomy between hosts of high- and low-excitation radio galaxies, and that it is the low luminosity radio sources that drive the AGN activity at $z<0.2$ \cite[e.g.][]{2011MNRAS.413.2815S}. A number of authors attempted to implement AGN feedback into galaxy evolution models \cite[e.g.][]{2009ApJ...699..525S}, but deep radio-loud AGN samples of wide range of luminosity densities and at $z>0.5$ are required to validate the models.

\item What is the kinetic luminosity function (KLF) for AGN? 

The radio luminosity density is the detectable signature of a radio-loud AGN, but, as we will discuss in this paper, this bears only a weak relationship to the intrinsic kinetic luminosity (jet power) of AGN. By providing multi-frequency, high-resolution images for large samples of radio-loud AGN, the SKA will give us the best possible chance to break the luminosity density/kinetic power degeneracy and therefore understand the power input by AGN to their host environments and supermassive black hole growth over cosmic time (for population studies see e.g. \citeauthor{2013AN....334..408K} \citeyear{2013AN....334..408K}, for a case study see \citeauthor{2012MNRAS.424.1774H} \citeyear{2012MNRAS.424.1774H}).

\item What drives the AGN evolution? Is the `radio-loud AGN' activity a singular phase for a galaxy, either long- or short-duration, or is it cyclic? What triggers the fuelling cycle? 

At present we know of a number of radio-loud AGN that show signatures of previous activity episodes \cite[][]{2000MNRAS.315..371S}, but it is still not clear whether all radio sources are re-triggered or only some fraction of them \cite[e.g.][]{2009BASI...37...63S}, or even what triggers radio activity. A number of authors have attempted to tackle this problem via both statistical population as well as case studies at low-redshifts \cite[for recent works see e.g.][]{2008MNRAS.388..625S,2012A&A...541A..62J,2014arXiv1411.2028K,2014arXiv1409.0566M}. However, deep radio-loud AGN samples of a wide range of luminosity densities at $z>0.5$ are required to extend these studies to higher redshifts \cite[e.g.][]{2014MNRAS.439..861K}.

\end{itemize}

In what follows we assume that continued progress in wide and deep optical/IR imaging and spectroscopic surveys will be such that it will be possible to {\it identify} a large fraction of the observed radio galaxy population and assign redshifts to the sources -- a precondition for any study of a population and its physics. Further discussions of radio-loud AGN hosts must involve a discussion of the expected optical survey coverage by the start of SKA science operations -- this is discussed elsewhere in this volume \cite[][]{AASKAAntoniadis,AASKABacon,AASKAKitching}. 

The chapter is composed as follows. In Section~\ref{sec:lifecycles} we present our current view of the lifecycle of a radio galaxy, and our current observational and theoretical understanding on the radio source evolution as it ages throughout its lifetime. In Sections~\ref{sec:birth} -- \ref{sec:death} we separately discuss each stage of a radio-loud AGN life, from a radio galaxy birth, through its mid-life, to its death; in each of those sections we consider the necessary SKA1 and SKA receivers for each of the radio source class (i.e. life phase) observations. In Section~\ref{sec:cosmo} we take a broader view on the radio-loud AGN populations, and discuss them in terms of AGN duty cycles, AGN unification and cosmological studies. A brief summary of the SKA elements for this study is given in Section~\ref{sec:conclusions}. We assume a flat Universe with the Hubble constant of $H_0 = 67$ km~s$^{-1}$~Mpc$^{-1}$, and $\Omega_{\rm M} = 0.685$ and $\Omega_{\Lambda} = 0.315$ \cite[][]{2013arXiv1303.5076P} throughout the paper.  

%---------------------------------------------------------------------------

\section{Lifecycles of radio-loud AGN}
\label{sec:lifecycles}

The typical timescales of AGN radio activity are estimated to be $\sim0.1$~Gyr \cite[e.g.][]{2000ApJ...544..671W,2012MNRAS.424.2028K,2012ApJ...756..116A}. Once the radio activity is triggered, the launched jet expands through the host galaxy and ambient medium until the jet supply ceases and the radio source slowly fades radiating away the remaining energy stored in radio lobes. A series of these events is what we refer to as a `lifecycle' of a radio source.

\subsection{How complete is our current picture?}
\label{sec:picture}

The observed populations of radio-loud AGN fall into reasonably well defined classes distinguished by radio morphology, luminosity density and physical size (the latter of which is generally interpreted to be proportional to age). The smallest size radio galaxies, the so-called Compact Symmetric Objects \cite[CSO, $<500$~pc;][]{1994ApJ...432L..87W}, Gigahertz Peaked Spectrum \cite[GPS, $<1$~kpc with turnover broadband radio spectra;][]{1970ApL.....6..201B,1990AA...233..379S,1991ApJ...380...66O} and Compact Steep Spectrum \cite[CSS, $<10$~kpc;][]{1982MNRAS.198..843P,1990AA...231..333F,1998PASP..110..493O} sources are compact radio sources completely embedded in the host galaxy. They are believed to be predominantly young, `start-up' or `baby radio-galaxies', approximately $10^3-10^5$~years old (Section~\ref{sec:birth}). These sources may be resolved at VLBI angular resolutions, where they often reveal morphologies similar to those of more extended, of the order of 100-kpc, sources \cite[e.g.][]{1995PNAS...9211447R,2000MNRAS.319..429S}.  At kiloparsec scales, the Fanaroff-Riley (\citeyear{1974MNRAS.167P..31F}) class I and II (FR~I, FR~II respectively) are distinguished  (Section~\ref{sec:midlife}). According to unification models \cite[][]{1995PASP..107..803U, 1996AJ....111...53O, 1997MNRAS.290L..17W, 1999MNRAS.304..160J}, these can be observed at various angles disguising themselves at times as, for example, core-dominated quasar and blazar sub-populations. FR~I and FR~II type sources are found to be typically $\sim10^7$~years old  (Section~\ref{sec:cosmo}).

A radio source is thought to evolve through these phases as it ages: from young, compact and luminous CSO/GPS and CSS sources, the jets of which strongly interact with dense inter-stellar medium (ISM) as they try to leave the host galaxy, to the large-scale FR~I and FR~II stage at which the relativistic jets extend into the inter-galactic (IGM) and intra-cluster medium (ICM). This paradigm was first proposed by \cite{1982A&A...106...21P} and \cite{1985MNRAS.215..463C}, and further refined by \cite{1995PNAS...9211447R}, \cite{1996ApJ...460..634R} and \cite{1999NewAR..43..675S,2000MNRAS.319..445S}, and is based on high resolution observations of young radio sources. With the recent observational advances we can now extend these evolutionary tracks to the very late stages of radio source evolution which include giant \cite[$>1$~Mpc;][]{2009ARep...53.1086K}, dying \cite[][]{2007A&A...470..875P}, and re-started radio galaxies \cite[][Section \ref{sec:death}]{2000MNRAS.315..371S}.

Whilst simple and appealing, this widely accepted evolutionary path may represent just one of many possible evolutionary tracks of radio-loud AGN -- perhaps the longest, main lifecycle path. Alternative paths may include sources that do not reach the giant phase stage, the FR~I/FR~II stage, or even the CSS stage \cite[e.g.][]{2003PASA...20...46M}. Recent observational evidence for such alternatives comes from the existence of so-called young faders \cite[][]{2004evn..conf...73K,2010MNRAS.408.2261K}, a class of compact, low radio luminosity density and small-scale CSS sources that resemble large-scale dying radio galaxies. It is still not clear what causes the radio activity to cease and why it may happen on a wide range of timescales, with some radio sources becoming long-lived giants and others dying in their infancy. For example, in the discussion on why only some FR~Is and FR~IIs evolve to Mpc scales, longer lifetimes, more powerful engines, or under-dense environments have been suggested as a solution, but no consensus has yet been reached \cite[][]{2006A&A...454...95M,2008MNRAS.385.1286J,2009ARep...53.1086K,2013AASP....3...42K}.

A true over-abundance of a young class of radio sources would advocate the view that some radio activity is indeed terminated prematurely. Our current radio source population counts suffer from well-recognised selection biases. Deep, sensitive and complete $N(z)$ measurements of each, well defined radio source type are required. However, selection criteria for the samples must ensure inclusion of all types of radio source at the same time to allow for the lifecycle analyses. Such deep measurements have also a potential to help us to investigate the weakly radio-emitting source population (weak-radio AGN), analyses of which are pivotal in the investigations of AGN duty cycles (Section~\ref{sec:cosmo}).

%---------------------------------------------------------------------------

\begin{figure}
\centering
\includegraphics[scale=0.64]{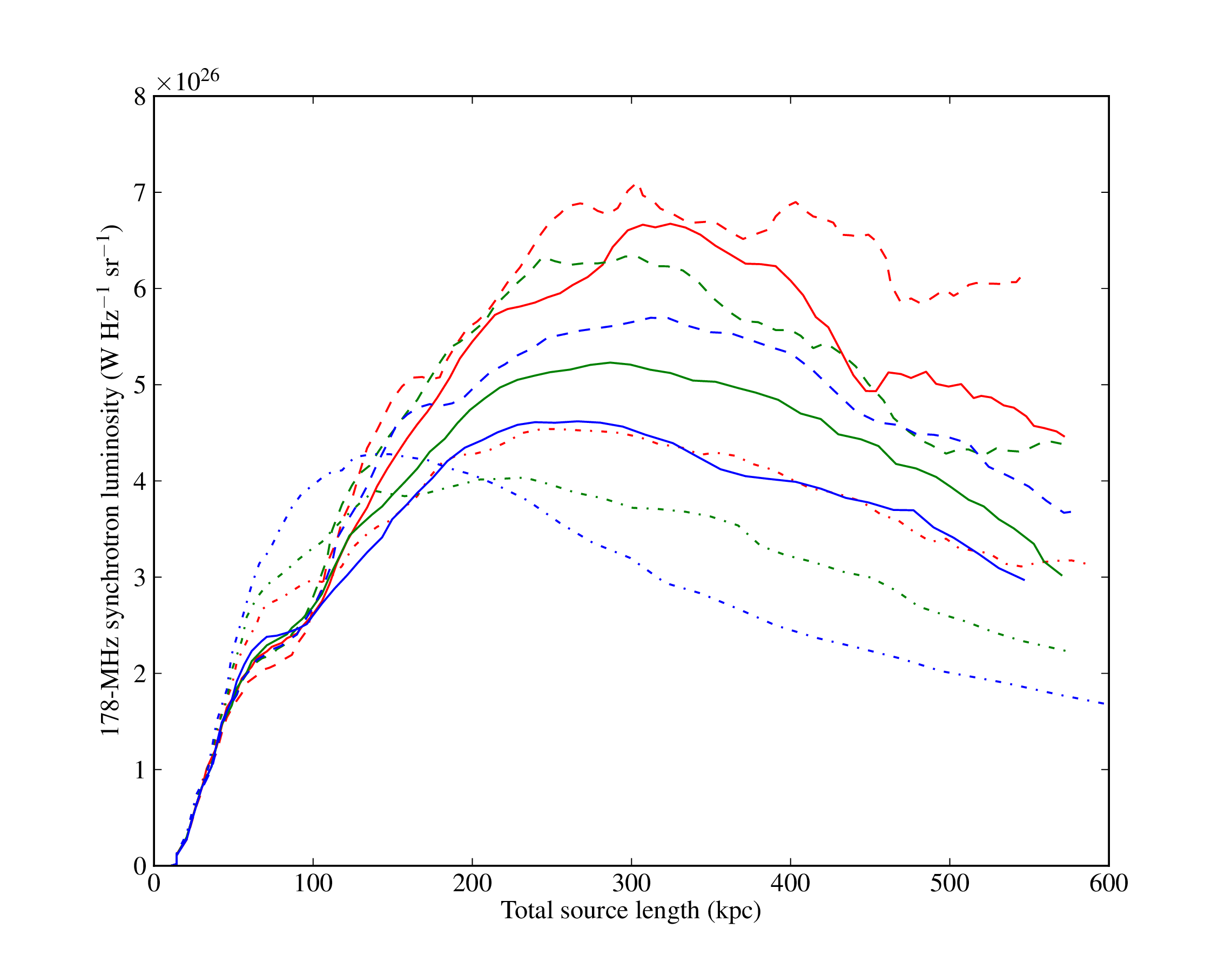}
\caption{The evolution of radio luminosity density as a function of source physical size as seen in 3D MHD simulations. The different coloured lines show the evolution of a radio source of the {\it same} jet power ($10^{38}$~W) in different plausible group/cluster environments, where dot-dashed lines represent poor environments, solid lines intermediate and dashed lines rich environments, and the colours denote the steepness of the density profile in the $\beta$-model \cite[][blue represent the steepest and red the flattest profile]{1962AJ.....67..471K}. Note the wide scatter in the luminosity densities at late times and the significant difference between the early luminosity densities and those in the `plateau' phase at a few hundred kpc. Standard estimates of jet kinetic power (based on scaling relations) would assign a wide range of different kinetic powers to the source simulated here, depending on the part of the lifecycle observed and the environment it inhabits. Figure adapted from \cite{2014MNRAS.443.1482H}.}
\label{rys:martin-synchr}
\end{figure}

%---------------------------------------------------------------------------

\subsection{Our theoretical understanding of radio-loud AGN evolution}
\label{sec:theory}

Building on the seminal work undertaken in the 1970s \cite[][]{1974MNRAS.169..395B,1974MNRAS.166..513S}, tremendous progress has been made especially in the past 20 years towards an analytical understanding of the physics and evolution of extragalactic radio sources. Semi-analytical approximations developed for classical double FR~II radio sources \cite[][]{1997MNRAS.286..215K,1997MNRAS.292..723K,1999AJ....117..677B}, are being extended to both young GPS/CSS \cite[][]{2000MNRAS.319..445S,2000MNRAS.319....8A,2006MNRAS.368.1404A,2012ApJ...760...77A,2014MNRAS.442.3469M} and dying radio galaxy stages \cite[][]{1994A&A...285...27K,2002MNRAS.336..649K}. The latter has been supported by observations of cavities in X-ray brightness maps of galaxy clusters \cite[considered to be signatures of radio source activity;][]{2004ApJ...607..800B,2007ARAA..45..117M}, as well as the discovery of so-called double-double (re-started) radio galaxies \cite[][]{2000MNRAS.315..371S}. The most recent developments include analytical modelling of re-started radio sources \cite[][]{2000MNRAS.315..381K,2011MNRAS.410..484B} that can account for multiple activity episodes. Very few radio sources with signatures of re-started radio activity have been observed to date -- the semi-analytical models are based on fewer than 30 known cases of re-started radio galaxies and on only one source with clear signatures of two episodes of previous activity \cite[][]{2007MNRAS.382.1019B,2011MNRAS.410..484B}. However, the existence of these sources and understanding of the physics involved is crucial in identifying the radio AGN activity (re)triggering processes and determining duty cycles of radio sources.  At the same time, numerical models involving realistic environments are now reaching the stage where they can be used to produce estimates of the evolution of integrated and resolved source properties that complement, and in some cases improve upon, those available from analytic modelling (e.g. Figure~\ref{rys:martin-synchr}).

On the other hand, FR~I sources are notoriously difficult to model analytically or numerically because of their complex, turbulent and often heavily disturbed radio structures. A number of attempts have been made \cite[][]{2003NewAR..47..577L, 2010ApJ...713..398L}, but no general analytical model exists that can predict properties of the FR~I class as a whole. This is particularly important, as FR~Is are thought to be much more numerous at low luminosity densities; it is possible that they will comprise the majority of the radio-loud AGN population at high redshift, low luminosity density radio samples; however, evidence for their high numbers is currently limited to a few studies in deep fields \cite[][]{2012MNRAS.421.3060S,2013MNRAS.436.1084M}. Numerical approaches to the problem of modelling FR~I radio galaxies have been so far focusing predominantly on dynamical evolution of their radio structures \cite[e.g.][]{2007MNRAS.382..527P,2014MNRAS.441.1488P}. Sensitive, deep SKA observations that will be available for a large number of such sources are required for further theoretical advances [detailed discussion on this subject is covered elsewhere in this volume; see \cite{AASKAAgudo,AASKALaing}]. 

However, analytical models investigating the plausible transition of FR~II sources into FR~Is have been developed \cite[][]{2011MNRAS.418.1138W,2014Turner}, and it is also thought that all radio sources start off with the FR~II morphology. Furthermore, we have also discovered a curious class of hybrid radio morphology objects, which show properties of both FR~I and FR~II sources \cite[][]{2000AA...363..507G,2014Kapinska}. This has direct implications for studies of the AGN host types, environments and their evolution across cosmic time, especially in terms of the observed FR dichotomy \cite[][]{2012AJ....144...85S}, and are already allowing us to statistically model young radio sources, FR~IIs and the FR transition populations. 

Clearly, observational advances drive our theoretical understanding of radio sources; only by combining both can we study the physical properties of radio-loud AGN that cannot be measured directly \cite[][]{1999AJ....117..677B,2012MNRAS.424.2028K}, but are crucial in the studies of galaxy and galaxy clusters evolution, and AGN feedback and activity.

\section{The birth of radio galaxies}
\label{sec:birth}

Young, compact radio sources are expected to be very numerous \cite[][]{1990AA...231..333F}, but are generally poorly studied in large samples. The observed complete, flux density limited parent samples often impose significant constraints on the population of radio galaxies that can be detected; most samples are biased towards middle-aged radio sources where the radio luminosity density is expected to be the highest for a given jet kinetic power (Figure~\ref{rys:martin-synchr}). There is also a strong bias of GPS/CSS sources to higher redshifts as compared to 100-kpc scale radio galaxies \cite[][]{2000MNRAS.319..445S}. In recent years, a number of faint GPS and CSS samples have been constructed probing lower intrinsic luminosities of these objects \cite[][]{1999NewAR..43..675S,2003A&A...402..171T,2010MNRAS.408.2261K}; but it is important to bear in mind that such samples are often constructed in a very different way than typical, complete samples of extragalactic radio sources, thus making it difficult to compare them to larger, and more evolved radio galaxies.

CSO/GPS and CSS sources are considered to be predominantly young radio galaxies, with typical ages of $10^3-10^5$~years \cite[][]{1998A&A...337...69O,1999A&A...345..769M,2003PASA...20...69P}. However, it is often difficult to distinguish truly young sources from objects whose expansion is `frustrated' by interaction with a dense ISM \cite[][]{1984AJ.....89....5V,1991ApJ...380...66O,2012ApJ...760...77A}. There is also increasing evidence that young double radio sources can have a substantial effect on the ISM of their hosts \cite[][]{2007ApJ...660..191C,2009MNRAS.395.1999C,2014MNRAS.439.1364H}; this seems to be also true for the small radio sources associated with canonically radio-quiet AGN \cite[][]{2012ApJ...758...95M}. 

Estimation of the ratio of young to older, large radio galaxies gives crucial constraints on the lifetime distribution of such objects (i.e. what fraction of them survive to the ages of $10^7$ years implied by dynamical and spectral ageing studies of 100-kpc scale radio galaxies), and hence on the accretion history of supermassive black holes that power them. It seems very plausible that there is substantial `infant mortality' in radio galaxies,  i.e. that many do not last long enough to reach the largest sizes and highest luminosity densities \cite[][]{2004evn..conf...73K,2010MNRAS.408.2261K,2012ApJ...760...77A,2014MNRAS.442.3469M}. This has been theoretically discussed by \cite{1997ApJ...487L.135R}, and fading CSS as well as GPS sources with re-started activity have also been observed \cite[][]{1990A&A...232...19B,2010MNRAS.408.2261K}. Deep and complete $N(z)$ measurements are crucial here to answer question whether all these small scale radio galaxies are progenitors of larger-scale FR~I and FR~II sources and to our overall understanding of the AGN lifecycles and the activity patterns of their central engines.

\subsection{Required SKA elements and the SKA surveys}

With the SKA1 baseline design \cite[][]{SKAbaselineDes}, survey depths will be such that we have a realistic chance of detecting the small-scale counterparts of all radio galaxies of even the lowest jet kinetic power  \cite[$10^{35}$~W;][]{2006MNRAS.370.1893H} out to $z\sim0.6$ using all-sky surveys with SKA1-SUR or SKA1-MID (Band 2, 1.4 GHz) and deep field observations with high frequency receivers on SKA1-MID (Band 4 and 5, 4~GHz and 9.2~GHz respectively). Assuming the jet kinetic power -- luminosity density scaling relation of \cite{1999MNRAS.309.1017W} holds, and scaling down the resulting luminosity density by a few orders of magnitude to account for evolution during the radio source growth (Figure~\ref{rys:martin-synchr}) we estimate luminosity densities of $\sim 10^{22}$~W~Hz$^{-1}$ for young radio sources which could be progenitors of the weakest FR~II radio galaxies. Given the numbers of FR~IIs, we would expect at least of order one of these young sources per square degree on the sky, and possibly many more if many FR~II-power jets turn off before they reach 100-kpc scales.

The main limitation of SKA1 for the study of CSS/GPS sources will be angular resolution; if we assume $0.05$~arcsec resolution at the highest frequencies (SKA1-MID Band 5), then at $z=0.3$ we will resolve only sources with linear sizes $>0.66$~kpc, and $>1.35$~kpc at higher redshifts (assuming source size at least $3\times$ the beam size). Lower frequencies (and especially SKA1-LOW) will not be useful for detailed radio morphology analyses. However, the high frequency capabilities will still allow an almost complete survey of the whole of the low- and mid-$z$ radio galaxy population down to the smallest sizes and lowest powers. Using a combination of angular resolution and in-band spectral information we will be able to distinguish young objects (steep spectrum, double lobe structure) from beamed, core-dominated systems (flat spectrum). The broad-band spectral coverage of SKA1-MID, -SUR and -LOW, especially when combined with each other, is crucial for selection of the GPS sources and investigation of their physics \cite[][]{2014Joe}. Early science can be carried out by aiming to be complete to some less ambitious combination of jet kinetic power lower limit and redshift upper limit, which will still give valuable insights into the properties of the lower-power sources.

All of these observations will provide a large number of sources too small for the SKA1 to resolve but which can be followed up by longer-baseline instruments such as the EVN, which are expected to provide an important complementary facility in the SKA era, unless the SKA VLBI facility is incorporated from the beginning of the telescope operations. Furthermore, detailed polarimetric studies of individual objects remain crucial to both understanding the relativistic jet--ISM interaction on parsec scales \cite[][]{2010MNRAS.402...87A,AASKAAgudo,AASKADehghan}, and establishing the fraction of young vs. frustrated CSO/GPS and CSS sources. 

In the final SKA stage we would expect to be able to see essentially every source with jet kinetic power $>10^{35}$~W, independent of its age except for sources less than a few hundred years old, out to $z=0.5-1.0$ (Bands 2 and 5), and all FR~II-power \cite[$>10^{36}$~W;][]{1991Natur.349..138R} start-up galaxies out to $z\sim3$, well into the regime where cosmological evolution of the radio source population becomes important. Assuming the angular resolution will be improved to at least $0.005$ arcsec, we will be able to resolve sources of linear sizes $>150$~pc at all redshifts.

%---------------------------------------------------------------------------

\begin{figure}
\centering
\includegraphics[scale=0.68]{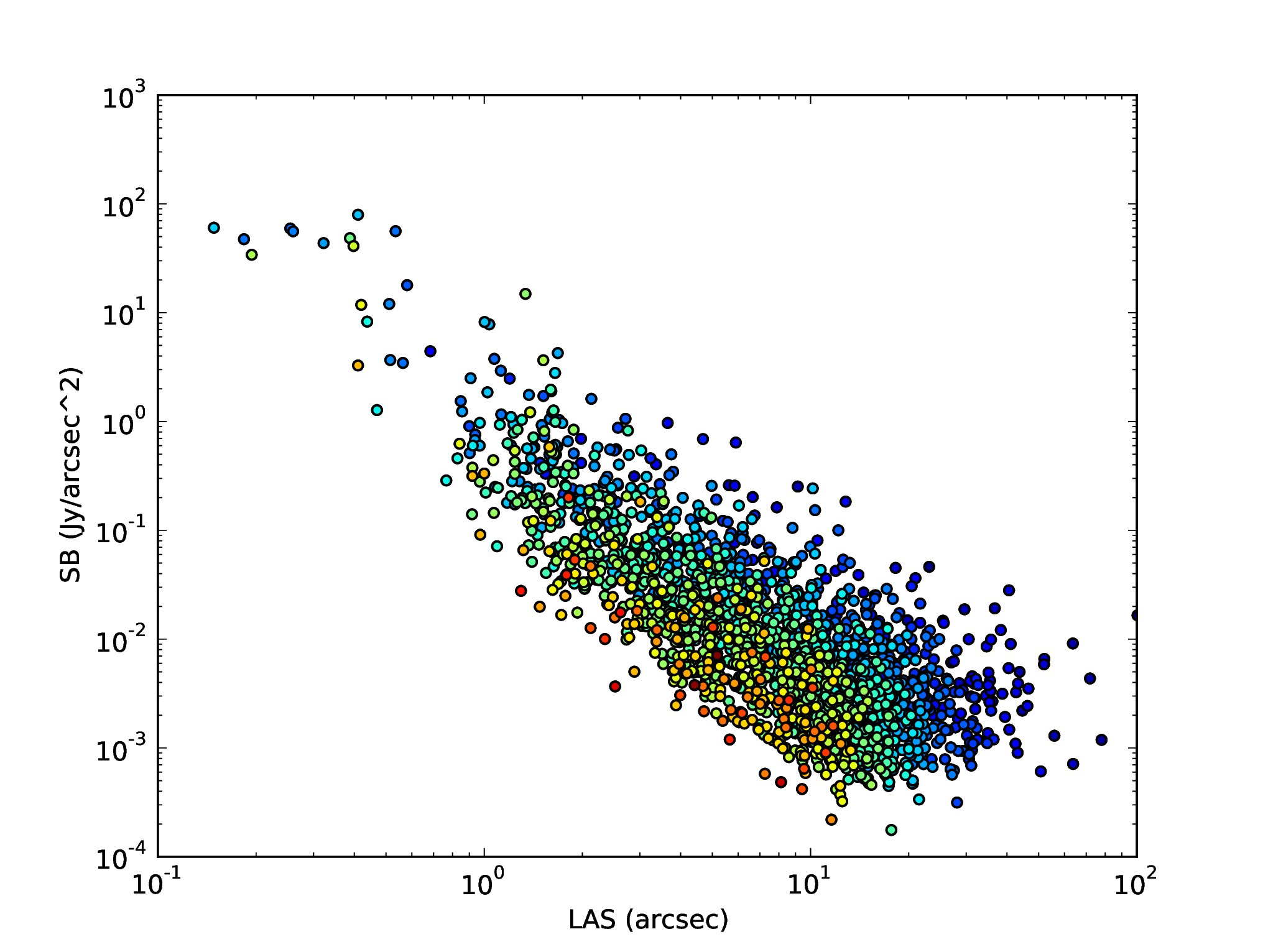}
\caption{Largest angular size (LAS) and 151 MHz surface brightness (SB) for FR~II radio galaxies from the SKADS SEX simulations \cite[][]{2008MNRAS.388.1335W}. Colour coding indicates redshift, from dark blue ($z=0$) to red ($z\simeq6$).}
\label{rys:martin-sizes}
\end{figure}

%---------------------------------------------------------------------------

\section{Radio galaxies in their mid-life: evolution, jet power and environmental impact}
\label{sec:midlife}

\subsection{Detailed physics of jets and radio lobes}

Large radio galaxies, with physical sizes of tens to hundreds of kpc, are the best-studied class of radio galaxy and have the best-understood effects on their ambient medium -- the hot phase of the IGM/ICM; these are the sources generally thought to be responsible for the `radio-mode feedback' that prevents the hot phase from cooling out onto the most massive galaxies \cite[][]{2004ApJ...607..800B,2007ARAA..45..117M}. As shown in Figure~\ref{rys:martin-synchr}, the brightness of the radio emission is expected to peak in this phase of a source's evolution, so these are the objects predominantly selected in large-scale flux density-limited surveys, and will be the easiest to detect and image with the SKA. 

Traditionally, radio galaxies detected in surveys have been characterized on the basis of an integrated flux density, and thus an estimate of radio luminosity density -- this can be directly used to construct RLFs of radio-loud AGN. However, for AGN feedback or radio source studies what one really wants to know is the jet kinetic power -- the instantaneous or time-averaged rate at which radio-loud AGN are transferring energy to their environments. The relationship between the observed radio emission and the jet kinetic power is a long-standing problem \cite[][see also Section~\ref{sec:theory}]{1991Natur.349..138R,1999MNRAS.309.1017W,2012MNRAS.424.2028K} and it is increasingly clear that the answer is not expected to be simple. At low radio luminosities, a substantial scatter in the relation between radio luminosity density and jet kinetic power is observed \cite[][]{2010ApJ...720.1066C,2013ApJ...767...12G}; this is expected on theoretical grounds, since the radio luminosity density should be a function of the jet kinetic power, the age of the source \cite[][]{1997MNRAS.292..723K}, radio morphology, and, crucially, the source's environment \cite[][]{1996MNRAS.283L..45B,2013MNRAS.430..174H} which we know to differ widely from source to source even at a fixed redshift \cite[][]{2013ApJ...770..136I}. In the case of weakly radio-emitting AGN, we can assess the interaction with their host environments just by measuring a luminosity density in some band and applying a bolometric correction. The problems described above mean that, for radio-loud AGN, no comparable correction exists.

%----------------------------------------------------------------------------------------------

\begin{figure}
\centering
\includegraphics[scale=0.65]{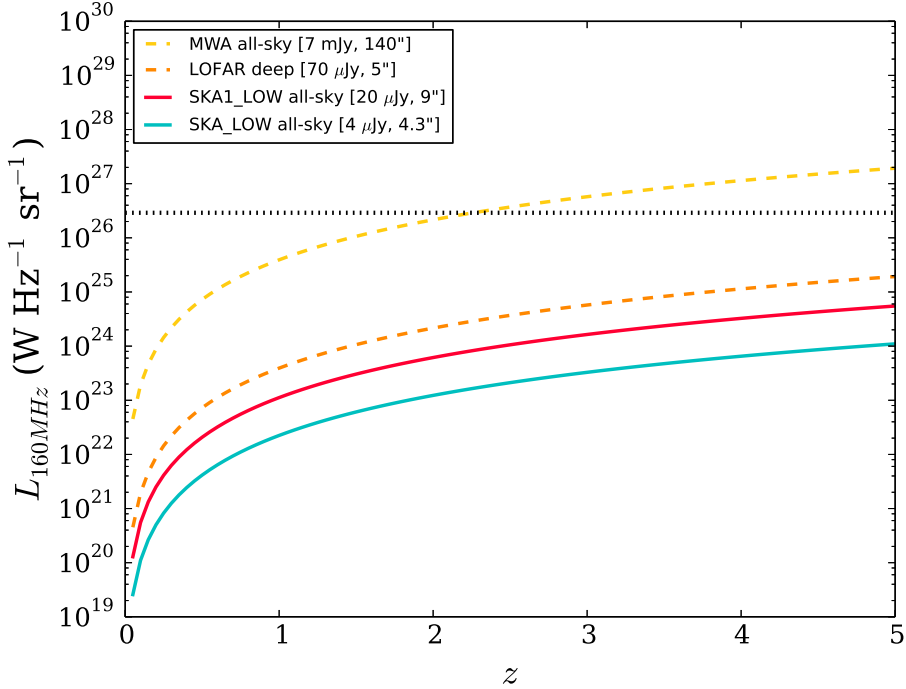}
\caption{$L-z$ limits of planned SKA1-LOW and SKA-LOW continuum all-sky surveys (solid lines; deep field observations are not feasible, confusion limited). We assume SKA-LOW will reach 90~km baselines (arm-core, $2\times$ SKA1-LOW) and angular resolution of $4.3$~arcsec. For comparison, the current SKA-LOW pathfinders and precursors are plotted (dashed lines): the ongoing MWA all-sky survey ($7$~mJy rms) and deep LOFAR Surveys observations (expected $70\mu$Jy rms for Tier 1). The horizontal line (dotted, black) marks the traditional FR~I/FR~II divide \cite[][]{1974MNRAS.167P..31F} scaled to 160~MHz (assuming $S\propto\nu^{-0.75}$). }
\label{fig:P-z-LOW}
\end{figure}

%---------------------------------------------------------------------------------------------

The jet kinetic power-environment-age degeneracy can be broken with observations that characterize not just the radio luminosity density but also the physical size, aspect ratio and spectral age of the source; in principle, this can allow not just the jet kinetic power but also the properties of the environment (possibly including an estimate of the heating rate) to be determined directly from radio observations. Spectral age measurements can be made by fitting to the  spatially resolved broad-band spectrum of the source \cite[][]{2013MNRAS.435.3353H}, and so to apply these techniques we need high spatial resolution (to image the lobes) as well as high spatial dynamic range (so that the largest scales are also well mapped). For this method to result in robust measurements, large samples of the order of hundred -- thousand sources are required.

\subsection{The role of SKA}

\begin{figure}
\centering
\includegraphics[scale=0.65]{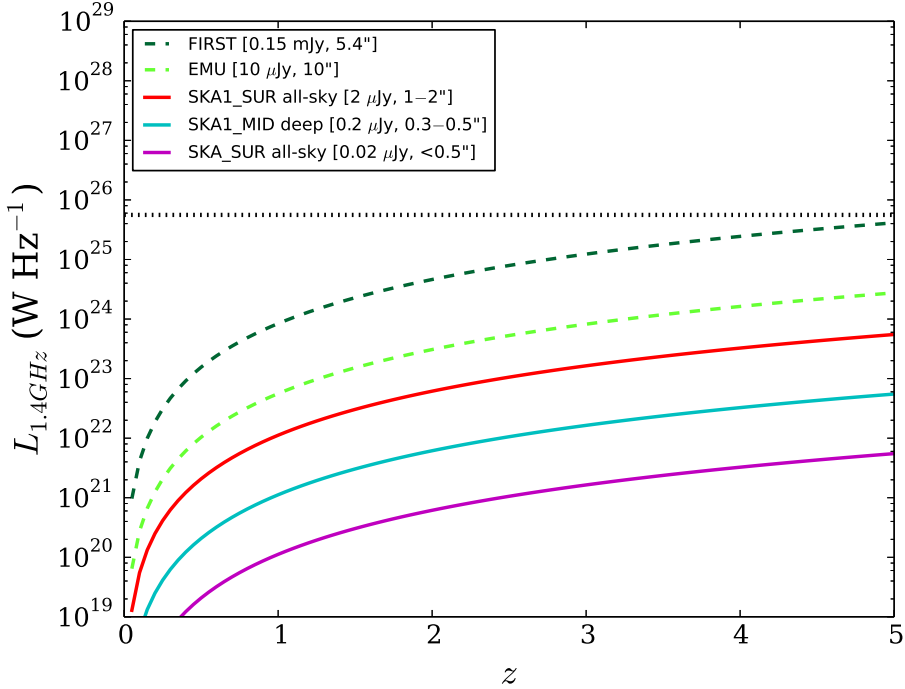}
\caption{$L-z$ limits of planned SKA1-SUR and SKA1-MID (Band 2) continuum all-sky and deep surveys. We assume SKA-SUR will reach $10\times$ the sensitivity of SKA1-MID. For comparison, the FIRST survey that have been serving us well over the past 20~years ($0.15$~mJy rms at 1.4~GHz) and the planned ASKAP-EMU all-sky survey (anticipated $10\mu$Jy) are plotted (dashed lines). The horizontal line (dotted, black) marks the traditional FR~I/FR~II divide \cite[][]{1974MNRAS.167P..31F} scaled to 1.4~GHz (assuming $S\propto\nu^{-0.75}$). }
\label{fig:P-z-SUR}
\end{figure}

To illustrate the potential of the SKA1 baseline design \cite[][]{SKAbaselineDes} in this area we consider the FR~II radio sources in the SKADS simulations of the extragalactic sky \cite[][]{2008MNRAS.388.1335W}. These simulations are probably not accurate enough to give reliable estimates for the properties of the young source population discussed in Section~\ref{sec:birth}, but should be adequate to consider the well-resolved FR~II population. We focus on the FR~IIs here since the dynamics and particle content of these `classical double' sources are (relatively) well understood \cite[][]{1997MNRAS.286..215K,2003NewAR..47..525C}. Figure~\ref{rys:martin-sizes} shows the mean 151 MHz surface brightness of all FR~IIs in the simulation as a function of source largest angular size; their surface brightness at 5 GHz is expected to be a factor 6 to 30 lower. With 0.05 arcsec beam (SKA1-MID Band 5) we can achieve modest levels of resolution (at least 5 beams across these large-scale sources) for most FR~IIs in the simulation, and do a lot better for the bulk of the population. While the $\mu$Jy sensitivity and exceptional $uv$-coverage of the projected surveys will allow the typical surface brightness of all but the faintest lobes to be imaged. Although at lower resolution, the SKA1-SUR in Band 2 (1.4~GHz) will be also very useful for this application as it will provide us with large, wide-field surveys of FR~Is and FR~IIs compiled within only 2 years on-sky time (Figure~\ref{fig:P-z-SUR}); we will still be able to resolve over half a population of these sources.

Good constraints on spectral age require broad-band measurements, ideally including low-frequency observations which would measure the un-aged energy spectrum of the electrons, the so-called injection index, which is known to vary significantly across the source population \cite[][]{2013MNRAS.436.1595K}. The reference design for SKA1-LOW \cite[][]{SKAbaselineDes}, with a beam at best a few arcsec, will not resolve the bulk of these sources, complicating the experiment. However, samples constructed from such low-frequency observations would be the base samples in our analyses of the AGN lifecycles and RLFs as they present the best probe of observing radio-loud AGN unbiased by the effects of relativistic beaming and orientation. At MHz frequencies, observations preferentially select jet and lobe emission due to its inherent injection spectral signature (i.e. very steeply rising spectra towards lower radio frequencies). The SKA1-LOW all-sky survey, completable within two years on-sky time, is expected to reach rms noise levels of $20\mu$Jy (confusion limited) at angular resolution of $\sim9$". This translates to luminosity densities of $5.7\times10^{23}$ W~Hz$^{-1}$ at $z=2$ (Figure~\ref{fig:P-z-LOW}) allowing us to detect the vast majority of CSS, FR~I and FR~II sub-populations out to very high redshifts ($z\sim4-5$ and beyond), but only at a modest spatial resolution of $20-75$~kpc depending on the redshift. 

While the SKA1 reference design \cite[][]{SKAbaselineDes} will allow many thousands of FR~II sources to be imaged to high redshift at high-to-moderate resolution, providing an essential test-bed for this technique, the final stage of the SKA   will be necessary to obtain well-resolved low-frequency images; this would require baselines of $>300$~km. Such baselines will allow us to resolve ($3-5\times$ beam size of 1.3~arcsec) over half of the population of FR~IIs at these low radio frequencies (Figure~\ref{rys:martin-sizes}). This would also extend the study to the much fainter, high-redshift FR~I population, and thus allow full analysis of the FR dichotomy at all epochs. Ideally, one would like to reach angular resolutions of 0.4~arcsec as this would allow us to resolve almost whole FR~II population; this however, would require baselines as long as $\sim1000$~km for SKA-LOW.

\section{Death, relics and activity re-triggering}
\label{sec:death}

\subsection{The population of relic radio galaxies}

The prominent features of radio sources (core, jets, hotspots) are fed by the continuous supply of energy from the active nucleus; once the jet activity stops, these features will disappear relatively quickly, and the lobe plasma will continue to expand and to cool via synchrotron and inverse-Compton losses, leaving a `relic' radio galaxy \cite[][]{1987MNRAS.227..695C}. During this fading phase very strong spectral evolution of the source occurs, with the high radio frequency part of the spectrum developing an ultra steep, exponential cut-off, and the spectral break shifting to lower radio frequencies.  Although every galaxy must go through this stage, only a handful examples of true dying radio sources is currently known \cite[][]{2007A&A...470..875P,2009ApJ...698L.163D}. Reasons for the rarity of such sources may be their low surface brightness and relatively short time they spend in the fading phase as compared to the average lifetime of a radio source; at GHz frequencies a source will fade away within $10^4-10^5$~years, while at MHz frequencies this may take $\sim10^7$~years. Identification of genuinely dying radio galaxies will give important information about lifetimes and duty cycles of extragalactic radio sources. Fading radio galaxies have also implications for AGN feedback since large amounts of the energy supplied by the jet remains stored in the lobes at the end of the active jet phase, and it remains an open question whether, and on what time and spatial scales, that energy is imparted to the ICM. 

As discussed in Section~\ref{sec:picture} the cessation of the jet energy supply seems to happen at any stage of radio source growth, and so we need to be searching for dying radio galaxies at all spatial scales, from pc to Mpc scales. Luminosity densities of lobes of fading radio galaxies are $0.3 - 40 \times 10^{23}$ W~Hz$^{-1}$ at 1.4~GHz \cite[][]{2007A&A...470..875P,2009ApJ...698L.163D}, while those of low-luminosity CSS sources, a significant number of which is believed to be in fading phase, to reach as low as $2\times 10^{23}$ W~Hz$^{-1}$ at 1.4~GHz \cite[][]{2010MNRAS.408.2261K}. With the currently available all-sky surveys \cite[e.g. FIRST, NVSS;][]{1995ApJ...450..559B,1998AJ....115.1693C} we are able to probe only the `tip of the iceberg' of this population, at relatively low redshifts ($>6\times 10^{23}$~W~Hz$^{-1}$ up to $z\sim0.3$ at 1.4~GHz) assuming the sources are not resolved out.

This is the biggest problem we are presently struggling with -- very few instruments can detect such low surface brightness sources, radio structures of which are often spread over multiple telescope's beams. One of the most spectacular recent examples of hidden imprints of previous activity (fading lobes) is 3C~452, which up to now was believed to be a classical FR~II radio galaxy \cite[][]{2013ApJ...765L..11S}. How many radio sources have previous activity signatures hidden in such a way? Clearly, there is a hidden world of secret lives of radio-loud AGN we are just starting to uncover. Presently advances and new discoveries are being already made with the existing and new facilities such as Murchison Wide-field Array \cite[MWA; e.g. ][]{2014Natasha}, Giant Metrewave Radio Telescope (GMRT) and Low Frequency Array (LOFAR).

%---------------------------------------------------------------------------------------------

%---------------------------------------------------------------------------------------------

\begin{figure}
\centering
\includegraphics[scale=0.65]{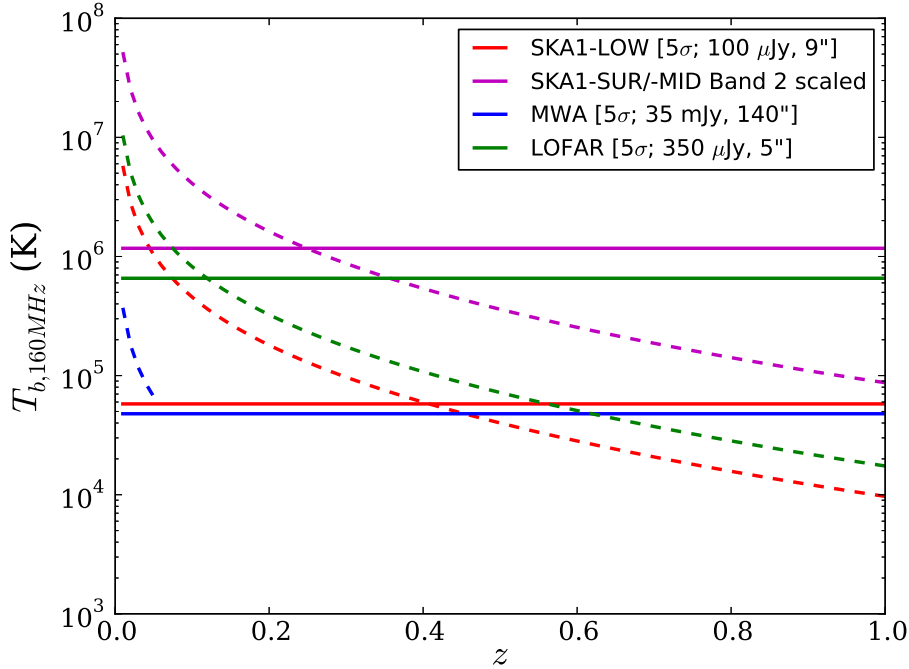}
\caption{Brightness temperature capabilities of SKA1-LOW and its precursors (MWA, LOFAR) for detecting fading radio source of 500~kpc physical size and total luminosity density $6\times10^{23}$ W~Hz$^{-1}$ at 160~MHz. Solid lines indicate the brightness temperature sensitivity  of each instrument for a given set of flux density limit and angular resolution; $5\sigma$ values are plotted. Dashed lines indicate the observed surface brightness temperature per beam of the assumed source, as a function of $z$, and are drawn only to a point where the angular size of the source is $>3\times$ the beam size of the instrument with which we want to detect the source. Both solid and dashed lines of the same colour refer to the same instrument. For a comparison, capabilities of SKA1-SUR/SKA1-MID Band 2 (1.4 GHz) in detecting such a source are plotted; predicted Band 2 all-sky survey sensitivity of $1\mu$Jy has been scaled to 160~MHz using an assumed steep spectral index of a fading radio source, $\alpha \sim 1.8$. For discussion see Section~\protect\ref{sec:death-ska}. Note that the surface brightness sensitivity of an instrument depends on both its sensitivity (flux density limit) and angular resolution. }
\label{fig:Tb-SKA}
\end{figure}

%---------------------------------------------------------------------------------------------

\subsection{What will SKA see?}
\label{sec:death-ska}

As an EoR driven instrument, SKA1-LOW will have tremendous surface brightness sensitivity capabilities, and so will be well adapted to searches for large, faint objects with little compact structure, although it may be difficult to identify them with their host galaxies. In principle, the number of large-scale fading radio galaxies should be comparable to the number of `alive' large-scale FR~Is and FR~IIs, but the fading radio galaxies will have much steeper spectra and considerably lower radio luminosity density; surface brightness sensitivity is one of the crucial aspects of the new radio telescopes if we want to obtain large samples of this rare class of radio source.  

To demonstrate capabilities of the SKA and its precursors, let us assume a fading radio source of an observed luminosity density $6\times10^{23}$ W~Hz$^{-1}$ at 160~MHz and 500~kpc total physical size. This luminosity density and physical size can be easily translated to a brightness temperature per telescope's beam and an angular size once a redshift is assigned. As shown in Figure~\ref{fig:Tb-SKA}, SKA1-LOW will be invaluable in searches for such radio sources, being able to detect and resolve them (at least with the modest $3\times$ beam size) up to $z\sim0.4$. Furthermore, we attempted to verify capabilities of higher frequency receivers of SKA for this application. Assuming a steep spectral index of the considered fading radio source, $\alpha\sim1.8$, its total luminosity density at 1.4~GHz would be $1.3\times10^{22}$ W~Hz$^{-1}$ spread over many telescope's beams, but the high sensitivity of SKA1-SUR/-MID will still be able to detect such a source out to $z\sim0.2$. Those higher frequency receivers will be able to detect less extreme sources at higher redshifts; for instance, a fading radio galaxy of a total luminosity density  $6\times10^{24}$ W~Hz$^{-1}$ at 1.4~GHz will be detectable out to $z\sim 1.6$, and although SKA1-LOW will detect sources of such luminosity density at 160~MHz $z\sim2$ and beyond, the former receivers will provide us with much higher resolution. A combination of both arrays may provide us with versatile range of images of dying radio sources.

\section{Evolution of the AGN population, radio-loud AGN unification and cosmology}
\label{sec:cosmo}

\subsection{Duty cycles of radio activity}
\label{sec:quiet}

Duty cycles, the relative times a radio source spends in its active and quiescent phases, is a crucial piece of information in our understanding of AGN lifecycles and their impact on the evolving Universe. To constrain these, we need to establish what causes radio sources to shut off and re-trigger, and whether all radio sources go through the re-triggering phase, or only a fraction of them. % {\bf (refs...)}.  
We need to construct representative RLFs not only of the radio-loud AGN in the midst of their activity (Sections~\ref{sec:birth}--\ref{sec:death}), but also of those in quiescence that are composed of the weakly radio-emitting AGN; these samples need to be well defined in terms of black hole accretion levels \cite[e.g.][]{2004NewAR..48.1157F,2008MNRAS.388.1011M}. 

It remains an open question as to whether there are underlying physical distinctions between the radio-loud/weak-radio AGN populations, or if there exists a continuum of radio activity which extends to the nuclear radio-quiescent galaxies, dominated by active or evolved star formation \cite[see e.g.][for recent analyses]{2011MNRAS.417..184B,2011ApJ...739L..29K,2012ApJ...754...12M}. The sensitive SKA continuum surveys will, for the first time, provide us with large samples of the largely unexplored weakly radio-emitting AGN ($10^{20}-10^{22}$ W~Hz$^{-1}$ at 1.4~GHz). During the SKA1 stage, with SKA1-SUR/SKA-MID (Band 2) all-sky surveys we will be able to detect the weak-radio AGN population down to luminosity densities of $5\times10^{21}$ W~Hz$^{-1}$ ($5\sigma$ detection) at $z=0.5$. With deep SKA1-SUR/SKA1-MID (Band 2) and SKA1-MID (Band 5) surveys we will go much deeper reaching $1\times 10^{21}$  W~Hz$^{-1}$ at $z=0.5$ and $4\times10^{19}$  W~Hz$^{-1}$ in the local Universe ($z\sim0.1$). SKA1-MID Band 5 receivers will also allow us to resolve these sources on linear scales of $0.3$~kpc at $z\sim0.5$. For detailed discussion on the physics of the weak-radio AGN and observations during the SKA era see \cite{AASKAOrienti}. 

\subsection{Radio-loud AGN unification}

The deep SKA surveys will sample all radio-loud AGN populations (Section~\ref{sec:intro}) with a subset of these being resolved in suitable detail to model their detailed internal physical processes. However, another view of the radio-loud AGN lifecycles is to view them in terms of populations that manifest themselves as quasars, radio galaxies, and blazars, depending on their orientation on the sky. Clearly, the detailed studies of these sources as described in this paper will improve our view of the radio-loud AGN populations and provide direct tests of simple unification models which extrapolate sparse RLF information with evolutionary scenarios to fit deep source count data \cite[e.g.][]{1996AJ....111...53O, 1997MNRAS.290L..17W,1999MNRAS.304..160J}. As discussed in Section~\ref{sec:theory} we model AGN lifecycles assuming that FR~Is and FR~IIs are physically distinct classes of radio source. This view is supported by simple unified models for radio-loud AGN where the FR~I and FR~II populations are the `parent' populations of many other observed classes of sources \cite[e.g.][]{1995PASP..107..803U, 1999MNRAS.304..160J}. In adopting these unified models, we can probe both the gross evolution of these populations and disentangle the effects of the AGN lifecycle, as well as map the contribution from weakly radio-emitting AGN and SF sources at least up to $z\sim0.5$. Angular resolutions of the order of arcsec are sufficient for these studies \cite[e.g.][]{2013MNRAS.436.1084M}.

Whilst testing the radio-loud AGN unification models we will obtain a simplified method to map the evolution of the radio-loud AGN population by exploiting the deep source counts, radio luminosity functions and populations under test. These analyses are best done from samples and source counts at low frequencies (<800 MHz); at these frequencies the sources are dominantly steep-spectrum and uncontaminated by relativistic beaming effects. For instance, following the methodology of e.g. \cite{1999MNRAS.304..160J} or \cite{2013MNRAS.436.1084M} we will be able to determine the evolution scenarios that fit best to the observed complete samples. These analyses may be used to provide insight into AGN duty cycles (Section~\ref{sec:quiet}), and the prevalence of certain types of sources such as GPS/CSS and dying radio galaxies. Using a range of deep counts at a range of observed frequencies (100 MHz -- few GHz) we can test evolutionary and lifecycle scenarios as well as explore the radio-loud / radio-quiet AGN divide.

\subsection{The SKA era precision cosmology}

Oneof the important aspects of the science described here is that we will identify different populations of AGN, and as presented in \cite{2014MNRAS.442.2511F} who showed that if one can separate out the various populations, which in turn sample the underlying dark-matter density distribution with a different bias, then the effects of cosmic variance may be overcome in determining the angular power spectrum. This is critical for studying the Universe on the largest scale \cite[][]{AASKACamera}. Powerful radio-loud AGN, provide a unique sampling of the underlying dark matter distribution as they are, generally, the most highly biased tracers of the density field, and are detected up to the highest redshifts ($z \sim 6$). Therefore, when combined with less biased tracers, e.g. star-forming galaxies \cite[][]{AASKAJarvis}, they may provide a unique way to understand the largest scales in the Universe, given that the ability to overcome the cosmic variance is dependent on the difference between the bias of the two populations under consideration. The key issue here is to separate AGN from star-forming galaxies, and the high-resolution of SKA1-MID can enable the separation of AGN and star-forming galaxies on either morphology for jet sources ($\gtrsim 0.5$~arcsec) or through pure brightness temperature measurements. Furthermore, knowledge of the $N(z)$ is also crucial and the detailed follow-up that would be carried out for this science case, in particular obtaining redshifts, will also be valuable for cosmological science.

\section{Concluding remarks}
\label{sec:conclusions}

A combination of the SKA arrays and their receivers at a wide range of frequencies and angular resolutions are necessary to address the science case discussed in this chapter. In particular, the SKA1-LOW array with its 300~MHz bandwidth, which we assumed here to be centered on 160~MHz, is an indispensable tool for searching for old, dying radio sources, and for the construction of complete, flux density and volume limited samples of radio-loud AGN unbiased by the effects of relativistic beaming.  SKA1-SUR and SKA1-MID Band 2 (centered on 1.4~GHz, with 800~MHz bandwidth) will provide us with deep, wide-field surveys at a reasonable resolution of $1-2$~arcsec. This will allow for morphological identification of large radio-loud AGN samples, as well as it will provide us with large weak-radio AGN samples in the local Universe. SKA1-MID Band 2, with angular resolution $3-5\times$ better ($0.3-0.5$~arcsec), can be used as a follow up for a number of deep fields. Finally, SKA1-MID with its Band 4 and Band 5 (centered on 4~GHz and 9.2~GHz, with 2.4~GHz and $2\times 2.5$~GHz bandwidths respectively) is particularly useful for high resolution (0.05~arcsec) imaging of jets and lobes of both young and evolved radio galaxies for detailed physics analyses. It will be also useful for distinguishing sources within our radio samples that are truly relativistically beamed. The SKA-VLBI facility, if incorporated from the beginning of the telescope operations, will be invaluable for detailed physics investigations of most of radio-loud AGN. 

A combination of these SKA arrays and frequency bands will provide us with broad-band radio spectra of the sources -- this is crucial for selecting certain types of radio source; e.g. GPS sources are distinguished by their turn-over spectra, while dying radio galaxies are extreme steep spectrum sources ($\alpha>1.8$). Furthermore, such broad-band spectra will allow for spectral age estimates of the radio sources so important for the radio-loud AGN physics and lifecycle studies. 

With such rich data sets we will be able to not only to investigate radio-loud AGN and their engine and model radio source detailed physics, but also to trace the AGN activity (triggering and feedback) up to the high-$z$ Universe and advance the galaxy formation and evolution models. Progress in these crucial science areas is currently hindered by the lack of wide-field, deep radio-loud AGN samples that extend to high-$z$, and so is limited only to the local Universe.

Advances are now being made with the SKA pathfinders and precursors, such as MWA, LOFAR, MeerKAT and ASKAP. SKA1  will be much faster than any of these precursors and pathfinders, completing all-sky surveys ($3\pi$~sr) within only 2~years on-sky time. It is, however, the full SKA that may revolutionise our understanding of the radio-loud AGN lifecycles and physics, by reaching the unexplored flux density depths of the radio sky. 

%---------------------------------------------------------------------------

\section*{Acknowledgements}
\noindent
ADK acknowledges financial support from the Australian Research Council Centre of Excellence for All-sky Astrophysics (CAASTRO), through project number CE110001020. CAJ is a West Australian Premier's Fellow and acknowledges support from the WA Government Department of Premier and Cabinet and from the Curtin University. TA acknowledges financial support from the China Ministry of Science and Technology (grant no. 2013CB837900). The authors thank the referee for detailed comments on the manuscript which improved completeness of this chapter.

%\bibliographystyle{mn2e}
%\bibliographystyle{apj-short-etal}
%\bibliographystyle{apj}
%\bibliography{mybib}

\end{document}